\documentclass[conference]{IEEEtran}
\IEEEoverridecommandlockouts

\usepackage{balance}
\usepackage{cite}
\usepackage{amsmath,amssymb,amsfonts,mathtools,amsthm}
\usepackage{algorithmic}
\usepackage{graphicx}
\usepackage{textcomp}
\usepackage{xpatch}
\usepackage{xcolor}
\usepackage{pgfplots,tikzscale,tikz}
\usepackage[utf8]{inputenc}
\usepackage{soul}
\DeclareMathOperator{\diff}{d}
\renewcommand{\baselinestretch}{0.986}

\newtheorem{remark}{Remark}

\makeatletter
\xpatchcmd{\@thm}{\thm@headpunct{.}}{\thm@headpunct{}}{}{}
\makeatother

\newcommand\drawline[1][black]{%
  \raisebox{2pt}{%
    \tikz \draw[#1,line cap=butt] (0pt,0pt) -- (10pt,0pt);%
  }%
}

\definecolor{mycolor1}{rgb}{0.00000,0.44700,0.74100}%
\definecolor{mycolor2}{rgb}{0.92900,0.69400,0.12500}%
\definecolor{mycolor3}{rgb}{0.46667,0.67451,0.18824}%

\begin{document}

\title{Dynamic droop specifications for Grid-Forming Inverter-Based Resources}


\author{Jennifer T. Bui, Dominic Gro{\ss}, Deepak Ramasubramanian
\thanks{J. T. Bui and D. Gro{\ss} are with the Department of Electrical and Computer Engineering at the University of Wisconsin-Madison, Madison, WI, USA (e-mail: \texttt{\{jtbui,dominic.gross\}@wisc.edu}). D. Ramasubramanian is with the Electric Power Research Institute, Knoxville, TN, USA (e-mail: \texttt{dramasubramanian@epri.com}). 

This material is based upon work supported by the U.S. Department of Energy's Office of Critical Minerals and Energy innovation under the Integrated Energy Systems Office under Award Number 38637. The views expressed herein do not necessarily represent the views of the U.S. Department of Energy or the United States Government.}
}

\maketitle

\begin{abstract}
The large-scale retirement of synchronous generators requires additional capabilities from inverter-based resources (IBRs) to ensure the stability and reliability of power grids.  With the heterogeneous controls of IBRs, it is especially important to understand their behavior on the grid. This work proposes a simple data-enabled dynamic model to capture the small-signal dynamics of IBRs and formulate specifications for grid-forming (GFM) IBRs. The dynamic droop model is complementary to well-studied impedance models and extends the common definition of steady-state droop coefficients to dynamic droop coefficients that fully characterize the IBR small-signal response below the nominal line frequency (e.g., subsynchronous oscillations). We propose bounds on the gain and phase of the dynamic droop coefficients to encode minimum requirements for GFM IBRs to promote interoperability and minimize adverse interactions. The resulting specifications also provide some insights into the much-debated question of how to certify an IBR as GFM. Moreover, we also provide dynamic droop specifications for frequency control ancillary services that, e.g., clarify and generalize the notion of an IBR inertia response. Finally, common grid-following (GFL) and GFM controls as well as original equipment manufacturer (OEM) models are used to illustrate the results and showcase the use of dynamic droop coefficients as a tool to screen IBR dynamics for potential adverse interactions.
\end{abstract}


\section{Introduction}

A major transition in dynamics, control, and operation of electric power systems is the increasing integration of power-electronic converters that interface renewable generation, energy storage systems, high voltage direct current (HVDC) transmission, as well as large loads. Notably, replacing the functionality of conventional bulk synchronous generators (SGs) with so-called inverter-based resources (IBRs) and energy storage significantly changes power system dynamics and challenges standard analysis, control, and operating paradigms and jeopardizes stability and reliability of bulk power systems~\cite{MDH+18}. On timescales of milliseconds to seconds, the associated challenges have often been framed around the loss of reliable frequency control and rotational inertia provided by synchronous generators. However, challenges such as subsynchronous oscllations, loss of voltage control in parts of a power system, and reduced system strength have arguably had a more significant impact in power systems in the last years. In particular, IBR dynamics can widely vary and interoperability of heterogeneous IBRs that interact through large-scale bulk power systems is a main concern.

Today, most IBRs use so-called grid-following (GFL) control that relies on a phase-locked loop (PLL) or zero-crossing detection for grid synchronization and assumes that a stable ac voltage waveform (i.e., frequency and magnitude) at the point of interconnection (i.e., relies on the presence of SGs). While GFL power converters can provide some (slow) ancillary services (e.g., primary frequency
control), dynamic stability of power systems can rapidly deteriorate as the share of GFL resources increases~\cite{MBP+2019}. To address this challenge, grid-forming (GFM) converters aim to impose self-synchronizing ac voltage dynamics at their terminal. While GFM controls such as droop control~\cite{CDA93}, virtual synchronous machine control~\cite{DSF2015}, and dispatchable virtual oscillator control~\cite{GCB+2019} are well established, no rigorous and verifiable  definition of GFM behavior exists. For example, GFM behavior is widely defined as an ac voltage source that maintains a nearly constant internal voltage phasor (i.e., phase angle and magnitude) in the sub-transient time frame. While useful, this specification lacks precision and cannot be directly tested.

One of the earliest attempts defines a GFM IBR as a stiff voltage source~\cite{RLB2012}. However, such a definition does not capture the self-synchronizing features of the GFM dynamics. Other definitions hinge explicitly on the specifics of the control implementation (i.e., the absence of a PLL). However, in practice specifications are required that are agnostic to the underlying power electronics and control architecture. A first attempt towards a definition that can be verified solely with input-output data has been made in~\cite{DDP2019}. In particular,~\cite{DDP2019} considers an IBR connected to an ac voltage source that, in abstraction, models a grid that is subject to frequency disturbances. To delineate between GFL and GFM control, the gain of the transfer function from ac voltage (i.e., grid) frequency disturbances to the IBR bus frequency is used as a metric. The results indicate that GFM IBRs show improved rejection of ''grid'' frequency disturbances at the IBR bus. While insightful, the underlying model and setup does not fully characterize the IBR small-signal dynamics and only captures a subset of IBR grid support functions. Recently, a similar setup has been used to bound the transfer functions from grid frequency and voltage perturbations to IBR active and reactive power injection~\cite{ESIG2025GFM}. While providing additional insights, the results largely focus on GFM IBR resembling a stiff voltage source behind an impedance between $4$~Hz and $40$~Hz that only represent a subset of GFM IBR implementations and functional capabilities.

Our main contribution is to extend the common definition of steady-state droop coefficients to so-called dynamic droop coefficients that fully characterize the IBR small-signal response below the nominal line frequency. The dynamic droop model is complementary to well-studied impedance models that have insightful and straightforward interpretation at or above the nominal line frequency~\cite{HBL2007,WWZ+2025,ANG+2025}. We develop bounds on the gain and phase of the dynamic droop coefficients to encode minimum requirements for GFM IBRs to promote interpretation and minimize adverse interactions. The resulting specifications provide some insights into the much-debated question of how to formally define GFM features. This result is illustrated by comparing the dynamic droop coefficients for a typical GFL and GFM IBR. Specifically, we identify the capability of an IBR to either stay reasonably in phase with steady-state droop specifications or suppress high-frequency oscillations as the core feature of GFM IBRs. Subsequently, we provide dynamic droop specifications for frequency control ancillary services that, e.g., clarify and generalize the notion of an IBR inertia response. Finally, original equipment manufacturer (OEM) models are used to illustrate the results and showcase the use of dynamic droop coefficients as a tool to screen IBR dynamics for potential adverse interactions.

\section{Dynamic Droop Coefficients}

\subsection{Dynamic Droop Coefficients}
We model the small-signal dynamics of an IBR using the frequency domain model
\begin{align*}
\!\begin{bmatrix} \Delta \omega (j 2 \pi f_p) \\ \Delta V (j 2 \pi f_p)\end{bmatrix}\! &=\! -\!\begin{bmatrix}
m_P (j 2 \pi f_p)\!\!\! & \xi_Q (j 2 \pi f_p)\\
\xi_P (j 2 \pi f_p)\!\!\! & m_Q (j 2 \pi f_p)
\end{bmatrix}\!\!
\begin{bmatrix} \Delta P (j 2 \pi f_p) \\ \Delta Q (j 2 \pi f_p)\end{bmatrix}\!,
\end{align*}
where $j$ is the imaginary unit, $\Delta \omega$, $\Delta V$, $\Delta P$, and $\Delta Q$ denote deviations of the AC voltage frequency, AC voltage magnitude, active power, and reactive power from their respective setpoints. This dynamic droop model has a straightforward interpretation and low dependence on the operating point for common GFM controls operating away from the IBR limits and frequencies $f_p\in\mathbb{R}_{>0}$ below the nominal system frequency $f_0 = \omega_0/(2\pi)$. A key advantage is that the dynamic droop model can readily be obtained from hardware experiments or simulations of black box models.

The inputs, outputs, and sign convention of the dynamic droop model are deliberately chosen to characterize the impact the IBR has on the grid and for ease of interpretation. In particular, the dynamic droop model fully characterizes the small-signal dynamics of an IBR by extending the common steady-state droop coefficients, denoted by $m^\star_P \in \mathbb{R}_{>0}$ and $m^\star_Q  \in \mathbb{R}_{>0}$ in this work, to frequency dependent droop coefficients $m_P (j  2 \pi f_p)$ and $m_Q (j 2 \pi f_p)$ that can be understood as the droop response to active and reactive power oscillations at a given frequency $f_p$. For instance, $m_P (j 2 \pi f_p ) \in \mathbb{C}$ models the gain and phase shift of the response of the IBR frequency to an active power oscillation with frequency $f_p$. Notably, if an IBR provides steady-state droop, then $\lim_{f_p \to 0} m_P (j 2 \pi f_p) = m^\star_P$ and $\lim_{f_p \to 0} m_Q (j 2 \pi f_p) =  m^\star_Q$.

Notably, for small $f_p$ (e.g., below $5$~Hz to $10$~Hz), the magnitude $| \xi_Q(j 2 \pi f_p) |$ and $| \xi_P(j 2 \pi f_p) |$ of the cross coupling terms are often negligible. Thus, the dynamics on timescales related to certain subsynchronous oscillations can typically be well approximated by a simplified model that only considers $m_P (j 2 \pi f_p)$ and $m_Q (j 2 \pi f_p)$. For brevity and clarity of the presentation, we will neglect $\xi_Q(j 2 \pi f_p)$ and $\xi_P(j 2 \pi f_p)$ in the remainder of this work. We emphasize that this assumption merely concerns the IBR transfer function and does not imply assuming decoupled power flow. 

\begin{remark}[\textbf{Off-diagonal droop coefficients}]\label{rem:offdiagonal}
The development of specifications considering the coupling between active power and voltage magnitude (i.e., $\xi_P(j 2 \pi f_p)$) and reactive power and frequency (i.e., $\xi_Q(j 2 \pi f_p)$), is out of the scope of this paper. At present, methods developed in control theory that enable decentralized (i.e., bus-level) stability conditions that ensure stability of the overall interconnected system are limited to single-input single-output bus dynamics~\cite{PM2019,BG+2026}. Extending these results to multi-input multi-output bus dynamics to derive practical specifications accounting for the entire bus dynamics and coupled power flow is the subject of ongoing work.
\end{remark}

\subsection{Experimental Identification}\label{sec:method}
\begin{figure}[b!!]
    \centering
\includegraphics[width=1\columnwidth]{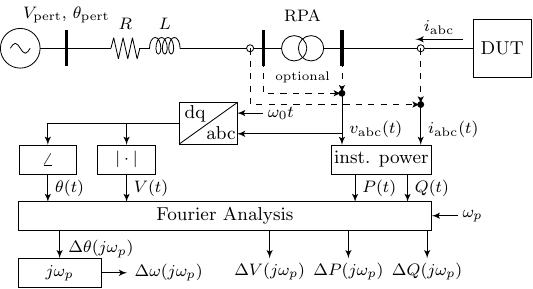}
\caption{Test setup and signal processing for experimental identification of dynamic droop coefficients. The AC voltage source injects frequency modulated perturbations into the device under test (DUT).\label{fig:testsetup}}
\end{figure}

To experimentally obtain the dynamic droop coefficients of an IBR, the setup shown in Fig.~\ref{fig:testsetup} can be used. In this work, an AC voltage source is used to perturb the IBR and recover the dynamic droop coefficients. Preliminary results indicate that using a current source to impose perturbations does not change the results for grid-forming IBRs. However, grid-following controls are expected to be unstable in such a configuration (i.e., dynamic droop coefficients cannot be obtained).
The overall approach consists of sinusoidal perturbations injected into the angle and magnitude of an AC voltage source with frequency $\omega_\text{pert} \in \mathbb{R}_{>0}$ and magnitude $V_\text{pert} \in \mathbb{R}_{>0}$ given by 
\begin{subequations}
\begin{align}
V_\text{pert} &\coloneqq V^\star + A_V \sin(2\pi f_p t),\\
\omega_\text{pert} &\coloneqq \omega_0 + A_\omega \sin(2\pi f_p t).
\end{align}
\end{subequations}
Here, $V^\star \in \mathbb{R}_{>0}$ denotes the nominal voltage and $A_V \in \mathbb{R}_{>0}$ and $A_\omega  \in \mathbb{R}_{>0}$ are the amplitudes of the sinusoidal disturbance injected at the voltage source. The voltage source angle is given by $\theta_\text{pert}= \int_{0}^{t} \omega_\text{pert} (\tau) \diff\tau$. The gains $A_V$ and $A_\omega$ need to be selected sufficiently large to ensure excitation of the IBR dynamics, yet small enough to ensure that the IBR remains well within its current, power, and modulation limits.

Simulations or experiments are conducted that separately perturb frequency (i.e., $A_\omega>0$, $A_V=0$) and magnitude (i.e., $A_V>0$, $A_\omega=0$) until signals have settled into a periodic steady state at frequency $f_p$. The three-phase voltage and current (flowing out of the IBR) waveforms $v_\text{abc} \in \mathbb{R}^3$ and $i_\text{abc} \in \mathbb{R}^3$ are captured at a reference point. Notably, the reference point does not have to coincide with the bus of the perturbation source and is solely determined by the location at which current and voltage measurements are taken. For example, if a transformer is present, the reference point could be before or after the transformer depending on whether the transformer is considered part of the IBR.

Next, the instantaneous active power $P$ and instantaneous reactive power $Q$ are computed. Moreover, the voltage phase angle $\theta$ and voltage magnitude $V$ are computed in dq frame with reference angle $\omega_0 t$. The amplitude and phase shift of the oscillations in the signals $V$, $\theta$, $P$, and $Q$ are recovered by computing their Fourier series coefficients corresponding to $f_p$. This allows representation of the magnitude and phase shift of the perturbations as complex numbers  $\Delta \theta$, $\Delta V$, $\Delta P$, and $\Delta Q$. Moreover, the magnitude and phase shift of the IBR frequency perturbation can be recovered as $\Delta \omega =j 2 \pi f_p \Delta \theta$. The complex perturbation signals can be collected in matrices
\begin{subequations}
\begin{align}
Y(j 2 \pi f_p) &= \begin{bmatrix}
\Delta \omega_\omega (j 2 \pi f_p) & \Delta \omega_V (j 2 \pi f_p)\\
\Delta V_\omega (j 2 \pi f_p) & \Delta V_V (j 2 \pi f_p)
\end{bmatrix},\\
U(j 2 \pi f_p) &= \begin{bmatrix}
\Delta P_\omega (j 2 \pi f_p) & \Delta P_V (j 2 \pi f_p)\\
\Delta Q_\omega (j 2 \pi f_p) & \Delta Q_V (j 2 \pi f_p)
\end{bmatrix},
\end{align}
\end{subequations}
where the subscripts $(\cdot)_\omega$ and $(\cdot)_V$ denote the data collected using frequency and magnitude perturbations, respectively. The dynamic droop model can then be immediately recovered by, e.g., computing the dynamic droop coefficients as
\begin{align*}
\begin{bmatrix} m_P(2\pi f_p) & \xi_Q (2\pi f_p) \\ \xi_P (2\pi f_p) & m_Q(2\pi f_p)\end{bmatrix} = -Y(j 2 \pi f_p) (U(j 2 \pi f_p))^{-1}
\end{align*} 
at any given perturbation frequency $f_p \in \mathbb{R}_{>0}$. 

The approach in Fig.~\ref{fig:testsetup} recovers the transfer function at the IBR terminal independently of the impedance of the perturbation source. In contrast, several frequency scan approaches reported in the literature determine the transfer function between the perturbation source (i.e., frequency and magnitude) and IBR power injection, i.e., inputs and outputs are not at the same bus. In this case, there is an increased risk of instability if the inductance $L$ between the perturbation source and the IBR is too small, while a sufficiently large impedance will reduce accuracy of the results.

\subsection{Example: Comparison of a GFL and a GFM IBR}
\begin{figure}[t!!]
    \centering
\includegraphics[width=1\columnwidth]{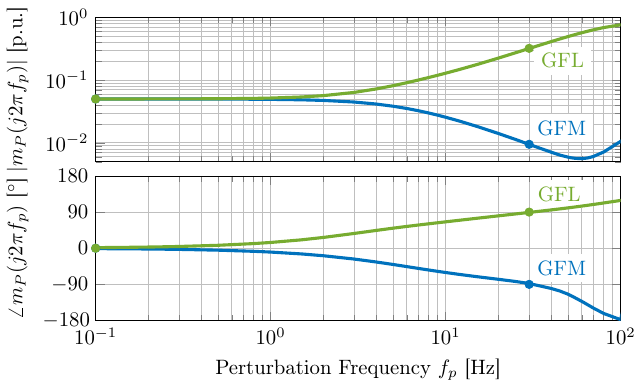}
\caption{Bode plots of $m_P(j2\pi f_p)$ for a prototypical GFM (\drawline[mycolor1, line width=1.3pt]) and GFL (\drawline[mycolor3, line width=1.3pt]) IBR obtained in EMT simulations using the approach presented in Sec.~\ref{sec:method}. Please see Fig.~\ref{fig:resp} for the time domain responses corresponding to the frequency response at $0.1$~Hz and $30$~Hz.\label{fig:bode}}
\end{figure}

The dynamic droop coefficients $m_P (j 2 \pi f_p)$ and $m_Q (j 2 \pi f_p)$ of a given IBR can be intuitively represented as a Bode plot with units of Hz instead of rad/s and gain in p.u. instead of decibel. Fig.~\ref{fig:bode} shows Bode plots of $m_P(j2\pi f_p)$ obtained in a detailed electromagnetic transient (EMT) simulation of GFM\footnote{Using vector PI voltage and current control in controller dq-frame.}  control and GFL\footnote{Using a synchronous-reference-frame PLL and vector PI current loop in controller dq-frame. The PLL frequency estimate and voltage magnitude to compute active and reactive power references.} control, both with 5\% steady-state $P\text{-}f$ and $Q\text{-}V$ droop. 
Notably, for $f_p \to 0$, the dynamic droop coefficient $m_P (j 2 \pi f_p)$ for GFL and GFM converge, highlighting that both can provide the same slow dynamic response~\cite{LGG2022}. However, the gain of $m_P (j 2 \pi f_p)$ decreases for GFM control as $f_p$ increases, i.e., the GFM IBR suppresses high-frequency oscillations. In contrast, for GFL control, the gain of $m_P (j 2 \pi f_p)$ increases as the frequency $f_p$ of active power perturbations increases resulting in reduced damping of frequency oscillations. 

\begin{figure}[t!]
    \centering
\includegraphics[width=1\columnwidth]{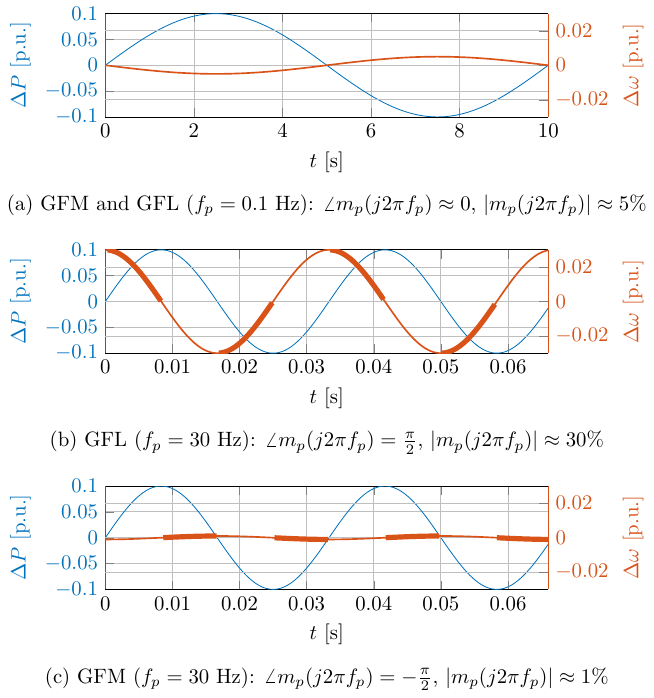}
\caption{Response of the IBR frequency (red) to a $0.1$~Hz and $30$~Hz active power oscillation. As the phase shift $\angle m_P(j 2 \pi f_p)$ increases, the response of the IBR frequency is increasingly out of phase of the steady-state droop specification $\Delta \omega = -m^\star_P \Delta P$ resulting in time instants at which $\Delta P$ and $\Delta \omega$ have the same sign (bold red), i.e., droop with the wrong sign. While the high gain of the GFL IBR will destabilize the system in this case, the low gain of the GFM IBR ensures stiff frequency control. \label{fig:resp}}
\end{figure}

To illustrate the differing dynamic droop behavior between GFL and GFM in time domain, Fig.~\ref{fig:resp} plots the frequency deviation $\Delta \omega$ in response to a $0.1$~Hz and $30$~Hz active power oscillation $\Delta p$.  In the first subplot, the GFL and GFM IBR responses to a $0.1$~Hz oscillation are the same.
As expected from the standard $P\text{-}f$ droop relationship, the frequency deviation $\Delta \omega$ has opposite sign to the power deviation $\Delta P$ if $m_P (j 2 \pi f_p )$ has zero phase shift. In contrast, at $f_p=30$~Hz both GFM and GFL control reach $\pm 90^\circ$ phase shift and the frequency deviation $\Delta \omega$ has the wrong sign for half the oscillation (bold in Fig.~\ref{fig:resp}). However, while the GFL IBR exhibits large frequency deviations, the GFM IBR exhibits small frequency deviations, i.e., stiff frequency control.

\subsection{Relation to Time-Domain and Impedance Specifications}
We emphasize that the frequency domain models and specifications proposed in this work are complementary to time-domain specifications and impedance specifications~\cite{HBL2007,WWZ+2025,ANG+2025} currently under discussion for GFM standards~\cite{UNIFI_Specs_V3,Fingrid_GESS_Spec_EN_2025,DE_4TSO_GFM_Reqs_2022,ENTSOE_GFM_PPM_Interim_2024}. In contrast to common time-domain specifications that characterize how an IBR responds to a specific network event, a key advantage of frequency domain specifications based on the dynamic droop coefficients is that they fully characterize the small-signal response and hence can help specify interoperability more broadly than specific scenarios considered in time-domain simulations. 

\begin{figure}[b!]
    \centering
\includegraphics[width=0.9\columnwidth]{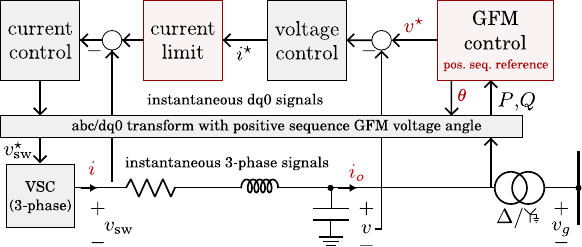}
\caption{Two-level voltage source converter with inner current and voltage control and outer GFM control.\label{fig:GFMarch}}
\end{figure}

To illustrate the complementary nature of the dynamic droop model and impedance models, consider the nonlinear dynamics of the GFM voltage source converter shown in Fig.~\ref{fig:GFMarch}. In particular, the GFM control is typically implemented as a linear system mapping active power $P$ and reactive power $Q$ to voltage phase angle $\theta$ and voltage magnitude $V$. For example, basic GFM droop control is given by
\begin{align}\label{eq:droop}
    \Delta \omega(s) = \frac{-m^\star_P}{Ts+1} \Delta P(s), \quad \Delta  V(s)= \frac{-m^\star_Q}{Ts+1} \Delta Q(s).
\end{align}
In contrast, the circuit dynamics and inner current and voltage control loops shown in Fig.~\ref{fig:GFMarch} are typically linear in the controller dq frame (i.e., with  angle $\theta$ provided by the outer GFM control), e.g., the filter inductor dynamics are given by a linear differential equation and the inner current and voltage control loops are typically proportional-integral controls.

Typically, stability of the converter controls is ensured by tuning the inner and outer control loops such that the dynamics of the converter filter circuit are fast enough and the dynamics of the outer GFM control loop are slow enough to not cause adverse interactions~\cite{SGC+2021}. Crucially, the fast dynamics of the IBR are largely determined by the circuit dynamics and inner loops that are linear in the controller dq frame and coordinates of impedance models~\cite{WWZ+2025} but nonlinear in coordinates of phase angle, voltage magnitude, and power. Conversely, the slow dynamics and interaction with other devices on the grid are largely determined by the outer GFM control loop that is linear in dynamic droop coordinates, but nonlinear in coordinates of impedance models.

Consequently, impedance models are a natural choice to accurately model and explain the fast dynamics of IBRs (i.e., beyond line frequency) but exhibit a significant dependence on the operating point at lower frequencies (see \cite[Fig.~10]{ANG+2025}). Conversely, the dynamic droop model is more insightful and shows less dependency on the operating point for slow dynamics (i.e., below line frequency) that are key drivers of subsynchronous oscillations. Similar arguments can be made for the SG stator dynamics (fast, linear in dq frame) and SG mechanics and turbine/governor (slow, largely linear in dynamic droop coordinates).

\section{Functional Capabilities of GFM IBRs and Stability Requirements}
Dynamics below $0.01$~Hz are largely governed by system-level and/or plant-level controls, and, in practice, no meaningful difference exists between GFM and GFL IBRs. On the other hand, due to the low-pass filter effects of transformers, dynamics beyond double line frequency do not significantly contribute to system-wide interaction and frequency synchronization of IBRs and/or SGs. Hence, we restrict our attention to the frequency range of $0.01$~Hz to $40$~Hz. We first focus on core functions that GFM IBRs can provide irrespective of resource capabilities. Subsequently, we discuss frequency control ancillary services that may require significant energy storage or power reserves (e.g., curtailment of resources).

\subsection{Mapping Functional Capabilities to Frequency Ranges}
In the remainder, the specifications for the dynamic droop coefficients are divided into four frequency regions. The region $f_p \in [0.01,f_p^\prime]$ parameterizes the quasi-steady-state response, the region $f_p \in [f_p^\prime,f_p^{\prime\prime}]$ parameterizes the transient droop response, and $f_p \in [f_p^{\prime\prime},f_p^{\prime\prime\prime} ]$ with $f_p^{\prime\prime\prime} \geq f_p^{\prime\prime}$ parameterizes the rate-of-change support (e.g., inertia). Finally, the region $f_p \geq f_p^{\prime\prime\prime}$ characterizes the response outside the bandwidth of GFM reference dynamics in which dynamics are predominantly governed by fast acting controls (e.g., current) and/or circuit dynamics (e.g., LCL filter). This region is crucial to distinguishing between GFM and GFL dynamics.

The specifications discussed in the remainder of this section were obtained from both functional/performance specifications~\cite[Sec.~6.1]{UNIFI_Specs_V3} and rigorous decentralized small-signal frequency stability conditions~\cite[Sec.~V]{BG+2026} that restrict the IBR transfer function to ensure stability of the overall interconnected system.  Notably, for $f_p \geq f_p^{\prime\prime\prime}$, the analytical stability conditions can be mapped to either low-gain or passivity conditions on $m_P (j2 \pi f_p)$. This flexibility is useful to accommodate~\cite[Sec.~V]{BG+2026} different GFM control architectures. For example, the low-gain specification may be more appropriate for GFM architectures with inner voltage and current control  (see Fig.~\ref{fig:GFMarch}) while the passivity specification may be more appropriate for GFM architectures without inner loops or virtual admittance control (see Sec.~\ref{sec:stiff}). 

\subsection{Frequency Control Ancillary Services}
Before focusing on specifications for GFM capabilities, we first discuss specifications for frequency control ancillary services that are solely related to the available IBR energy and power reserves. Notably, IBRs with significant energy storage (e.g., battery energy storage systems) or significant resource reserves (e.g., curtailed renewables) can provide frequency control services, such as steady-state droop, low-frequency oscillation damping, and an inertia response. These specifications are not strictly related to distinguishing between GFL and GFM control, i.e., GFL IBRs may be able to provide some frequency control ancillary services.

\begin{figure}[b!!]
    \centering
\includegraphics[width=1\columnwidth]{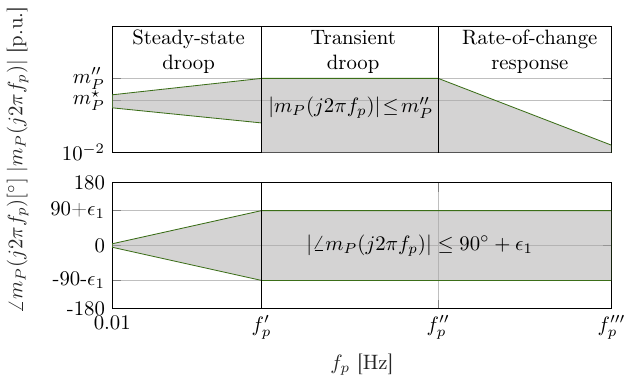}
\caption{Specification on $m_P( j 2 \pi f_p)$ for steady-state droop, transient droop, and rate-of-change response (i.e., inertia).\label{fig:anc}}
\end{figure}

Requirements on $m_P (j2 \pi f_p)$ that formalize provision of short-duration and long-duration frequency support are shown in Fig.~\ref{fig:anc}. Notably, as $f_p \to 0.01$~Hz, steady-state droop requires that $m_P (j2 \pi f_p)$ closely matches the provided specification, i.e., $|m_P (j2 \pi f_p)| \approx m^\star_P$ and $\angle m_P (j2 \pi f_p)| \approx 0$.
Between $f_p^\prime$ and $f_p^{\prime\prime}$ the gain $|m_P (j2 \pi f_p)|$ shall be bounded by the transient droop gain $m_P^{\prime\prime}$ to ensure sufficiently stiff frequency on transient timescales both from IBRs and SGs. Finally, between $f_p^{\prime\prime}$ and $f_p^{\prime\prime\prime}$ the gain is required to decrease ten times per ten times increase in $f_p$ to encode a rate-of-change (i.e., inertia) response, i.e., an increasingly stiff frequency. Notably, for a GFM VSM with inertia constant $H$ and steady-state droop $m^\star_p$ (i.e., \eqref{eq:droop} with $T=2H m^\star_p$) one can verify that $f_p^{\prime\prime} \approx 1/(4 \pi H m^\star_p)$. The transient upper bound $m_P^{\prime\prime}$ on the active power droop is assumed to be selected to enforce existing requirements such as those outlined in generator prequalification tests for primary frequency control. In other words, if the transient active power response to a frequency step is required to be within some percentage $\gamma$ of $\Delta P = -1/m^\star_P \Delta \omega$, then $m_P^{\prime\prime}=(1+\gamma) m^\star_P$ can be used to encode this requirement.

\subsection{Frequency Synchronization and High-Frequency Stiffness}
\begin{figure}[b!!]
    \centering
\includegraphics[width=1\columnwidth]{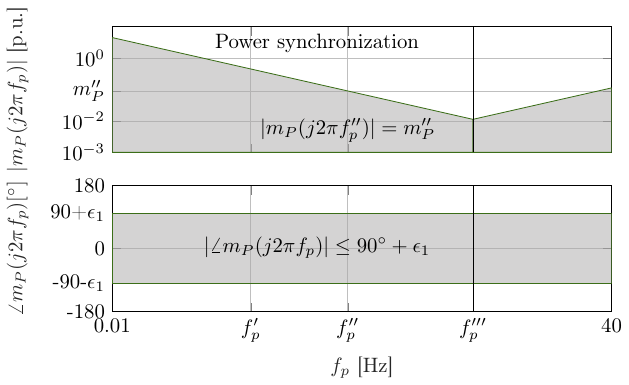}
\caption{Power synchronization specification for $m_P(j2 \pi f_p)$ and passive high-frequency voltage phasor stiffness.\label{fig:PFspec}}
\end{figure}
The dynamic droop specifications on $m_P(j2\pi f_p)$ encode minimum requirements for GFM IBRs aimed at ruling out adverse interactions. To this end, the specification shown in Fig.~\ref{fig:PFspec} allows for proportional-integral (PI) active power-frequency droop control that enables GFM synchronization without providing steady-state droop (e.g., GFM STATCOM) while meeting the transient droop gain specification, i.e., the upper bound on $|m_P(j 2 \pi f_p)|$ decreases ten times for a ten times increase in $f_p$ for $f_p \leq f_p^{\prime\prime\prime}$ and we require $|m_P(j 2 \pi f^{\prime\prime}_p) | = m^{\prime\prime}_P$. The phase is restricted to $|\angle m_P(j2 \pi f_p)| \leq 90°+\epsilon_1$ to prevent a droop response with incorrect sign (on average over one cycle of an oscillation), where a small tolerance $\epsilon_1$ is added to account for numerical errors. Beyond $f_p^{\prime\prime\prime}$, either one of two specifications can be used that trade off gain and phase conditions~\cite[Sec.~V-B]{BG+2026}.

The passive high-frequency specification shown in Fig.~\ref{fig:PFspec} allows for $|m_P(j 2 \pi f_p)|$ to increase ten times for a ten times increase in $f_p$ to account for the impact of output impedance of a GFM converter controlled as a voltage source behind an impedance. GFM controls without inner control loops or virtual admittance voltage controls typically fall into this category (see Sec.~\ref{sec:stiff}). 

Alternatively, arbitrary phase $\angle m_P(j 2 \pi f_p)$ beyond $f_p^{\prime\prime\prime}$ is permissible only if the gain is sufficiently low (i.e., $|m_P(j2 \pi f_p)|\leq m_{\text{low}}$) to minimize the impact of negative damping when the droop response is more than $90^\circ$ out of phase. GFM control architectures with proportional-integral (PI) current and voltage control (see Fig.~\ref{fig:GFMarch}) often fall into this "low gain" category.

\subsection{GFM Voltage Support}
Finally, we discuss specifications on the voltage response that are independent of the IBR energy and power reserves. In the lower frequency range, $m_Q(j2 \pi f_p)$ must closely match the steady-state droop specification, i.e., $|m_Q(j2 \pi f_p)| \approx m^\star_Q$ and $\angle m_Q(j2 \pi f_p) \approx 0$ as $f_p \to 0.01$~Hz. Next, between $f_p^\prime$ and $f_p^{\prime\prime\prime}$ the gain shall be bounded by the transient droop gain $m_Q^{\prime\prime}$ and the phase is restricted to $|\angle m_Q(j2 \pi f_p)| \leq 90°+\epsilon_1$ below $f_p^{\prime\prime\prime}$ to prevent a droop response with incorrect sign (on average over one cycle of an oscillation). If the reactive power response to a step in voltage is required to remain within a certain percentage of the steady-state response then $m_Q^{\prime\prime}$ should be selected within the same percentage of $m_Q^\star$. Beyond $f_p^{\prime\prime\prime}$, either a passivity or small gain condition can be used.

\begin{figure}[b!!]
    \centering
\includegraphics[width=1\columnwidth]{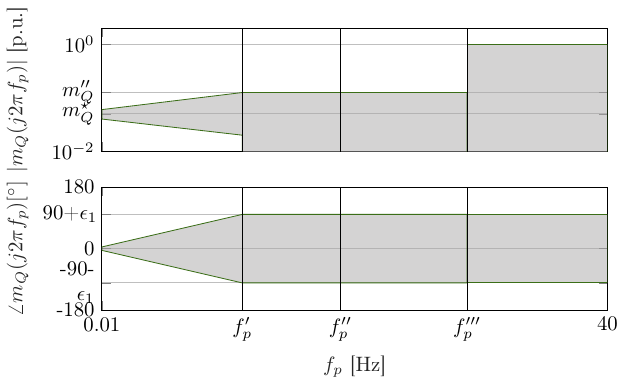}
\caption{Specification on $m_Q(j2 \pi f_p)$ for $Q\text{-}V$ droop and passive high-frequency voltage phasor stiffness.\label{fig:QVspec}}
\end{figure}

The passive high-frequency specification shown in Fig.~\ref{fig:QVspec} restricts the phase to $|\angle m_Q(j2 \pi f_p)| \leq 90°+\epsilon_1$ and the gain to $1$~p.u., which allows for the impact of the output impedance of a GFM converter controlled as a voltage source behind an impedance. Alternatively, arbitrary phase $\angle m_Q(j2 \pi f_p)$ beyond $f_p^{\prime\prime\prime}$ is permissible only if the gain is sufficiently low (i.e., $|m_Q(j2 \pi f_p)|\leq m_{\text{low}}$) to minimize the impact of negative damping when the droop response is more than $90^\circ$ out of phase. Preliminary results suggest that, for a Thevenin equivalent grid reactance $X_\text{g}$ in p.u., $m_\text{low} \approx f_0/f_p^{\prime\prime\prime} X_\text{g}$. The specifications on $m_Q(j2 \pi f_p)$ are illustrated Fig.~\ref{fig:QVspec} for the case of passive high-frequency dynamics.

\subsection{Parameter Ranges and Operating Conditions}
We envision that requirements on the frequencies $f^\prime_p$, $f^{\prime\prime}_p$, and $f^{\prime\prime\prime}$ are specified by a system operator based on prevailing timescales of frequency and voltage control in the system. In contrast, the IBR may need to comply with steady-state droop constants (e.g., $m^\star_P$) and transient droop bounds (e.g., $m^{\prime\prime}_P$) that are updated depending on operating and system conditions. To this end, compliance with the entire range of applicable droop constants and bounds should be established before commissioning using the method outlined in Sec.~\ref{sec:method}.

\section{Examples and Case studies}

\subsection{Stiff Voltage Source Behind an Impedance}\label{sec:stiff}
A widely discussed specification for GFM IBRs is to behave like a stiff voltage source behind an impedance. While this specification is in direct conflict with self-synchronization through droop, it provides useful insights at higher frequencies. Notably, if no inner control loops are used, the filter impedance typically dominates the response of the IBR for $f_p \geq f_p^{\prime\prime\prime}$. Specifically, when providing a rate-of-change response with cut-off frequency $f_p^{\prime\prime} \ll f_p^{\prime\prime\prime}$, the dynamics of the GFM IBR without inner controls will resemble a stiff voltage source for frequencies $f_p \in [f_p^{\prime\prime\prime},f_0]$. Next, consider the transfer function
\begin{align*}
\Delta P(s) = \frac{1}{X_\text{o}}\frac{\omega_0^2}{s^2+2 \omega_0 \rho s+\omega_0^2+\omega_0^2 \rho^2} \delta(s),
\end{align*}
from the voltage phase angle difference $\delta$ to active power $\Delta P$~\cite{G22}, with reactance $X_\text{o}$ in per unit and resistance-reactance ratio $\rho$. For a stiff voltage source behind a reactance, $m_P (s) = (s/\omega_0) X_\text{o}(s^2+2\omega_0 \rho s+\omega_0^2+\rho^2)/\omega_0^2$. For small $\rho$ i.e., $\rho<1/10$) this transfer function can be approximated by $m_P (j2 \pi f_p ) \approx j X_\text{o}  f_p/f_0$ in the frequency range $f_p \in [f_p^{\prime\prime\prime},f_0]$, in other words, a ten times increase in gain for every ten times increase in frequency and a phase of $90^\circ$. This high frequency asymptote can be observed as an upper bound in Fig.~\ref{fig:PFspec}. Notably, the asymptote shifts up if $X_\text{o}$ is increased.

Moreover, consider the transfer function
\begin{align*}
\Delta Q(s) = \frac{1}{X_\text{o}}\frac{\omega_0^2}{s^2+2 \omega_0 \rho s+\omega_0^2+\omega_0^2 \rho^2} \Delta V(s),
\end{align*}
from voltage magnitude difference $\Delta V$ to reactive power $\Delta Q$ flowing through the reactance~\cite{G22}. This results in the dynamic droop coefficient $m_Q (s) = X_\text{o}(s^2+2\omega_0 \rho s+\omega_0^2+\rho^2)/\omega_0^2$ for the response of a stiff voltage source behind a reactance. This reduces to $m_Q (s) \approx X_\text{o}$ for small $\rho$ in the frequency range of interest. Thus, the specification $|m_Q(j 2\pi f_p)| \leq 1$~p.u., requires that $X_\text{o} \leq 1$~p.u., precluding unreasonably high output reactance that would significantly degrade grid strength.

\subsection{Kaua`i Island Power System $18$-$20$~Hz Oscillations}
\begin{figure}[t!!]
    \centering
\includegraphics[width=1\columnwidth]{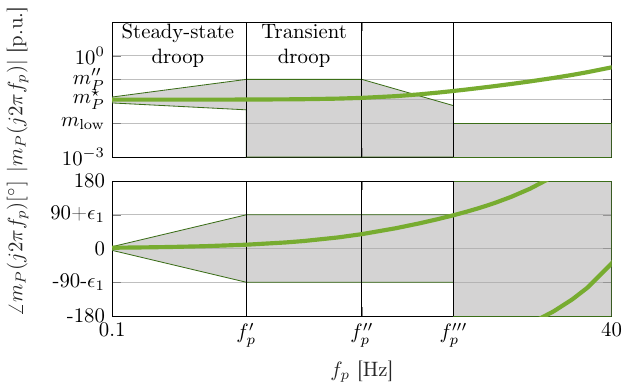}
\vspace{-2em}
\caption{Dynamic droop coefficient $m_P( j 2 \pi f_p)$ for a GFL (\drawline[mycolor3, line width=1.3pt]) IBR that exhibits $180^\circ$ phase shift at approximately $19$~Hz.\label{fig:GFLdel}}
\end{figure}

To showcase the use of dynamic droop coefficients as a screening tool, we consider an EMT model of a typical GFL IBR (see \cite[Fig.~4a]{RLB2012}) with $P\text{-}f$ droop that has been modified to replicate the $18$-$20$~Hz oscillation in the Kaua`i Island Power System in simulation~\cite{dong2023analysisnovember212021}. A Bode plot of the dynamic droop coefficient $m_P( j 2 \pi f_p)$ has been obtained using the method described in Section~\ref{sec:method} and is shown in Fig.~\ref{fig:GFLdel}. It can be seen that this GFL IBR delivers steady-state and transient droop, but does not provide rate-of-change support and fails to meet the high-frequency specifications for GFM IBRs. In particular, for $f_p > f_p^{\prime\prime\prime}$, we observe a phase shift beyond $90^\circ$ and high gain. Therefore, this IBR does not qualify as GFM. In addition, the phase shift is $180^\circ$ for $f_p \approx 19$~Hz, indicating negative damping (i.e., $P\text{-}f$ droop with flipped sign). Thus, it is not surprising that the dynamics of this GFL IBR have been identified as the source of a $18$-$20$~Hz oscillation.

\subsection{Battery Energy Storage System (OEM~1)}
\begin{figure}[b!!]
    \centering
        \vspace{-1em}
\includegraphics[width=1\columnwidth]{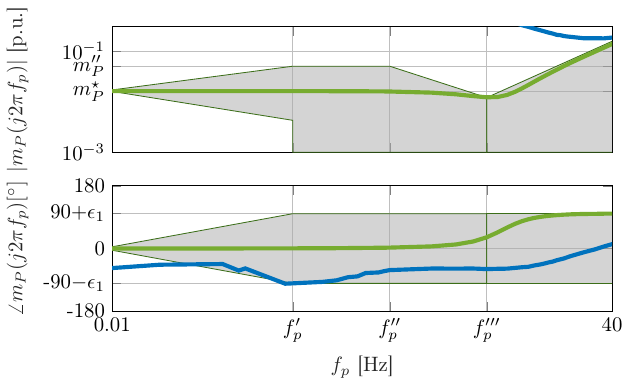}
\vspace{-2em}
\caption{Dynamic droop coefficient $m_P( j 2 \pi f_p)$ for a GFL (\drawline[mycolor3, line width=1.3pt]) and GFM (\drawline[mycolor1, line width=1.3pt]) configuration for the model provided by OEM~1.\label{fig:OEM1_mp}}
\vspace{0.5em}
\includegraphics[width=1\columnwidth]{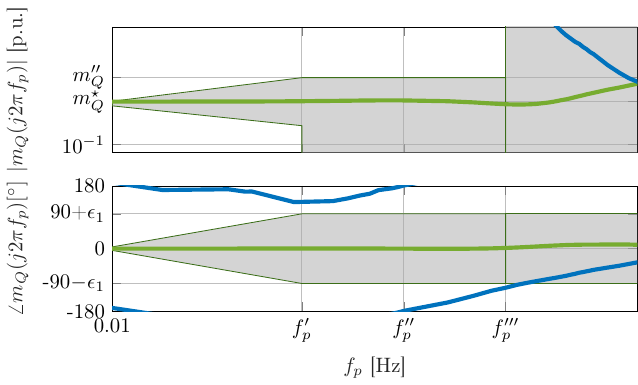}
\vspace{-2em}
\caption{Dynamic droop coefficient $m_Q( j 2 \pi f_p)$ for a GFL (\drawline[mycolor3, line width=1.3pt]) and GFM (\drawline[mycolor1, line width=1.3pt]) configuration for the model provided by OEM~1. \label{fig:OEM1_mq}}
\end{figure}
Next, we scan a model of a $3.4$~MVA, $600$~V GFL and GFM IBR provided by a vendor (OEM~1). In the GFM model, the steady-state droop is parameterized as $m_Q^\star=0.18$ and $m_P^\star=0.016$. Bode plots of the dynamic droop coefficients $m_P( j 2 \pi f_p)$ and $m_Q( j 2 \pi f_p)$ are shown in Fig.~\ref{fig:OEM1_mp} and  Fig.~\ref{fig:OEM1_mq}. The GFM model aligns with the proposed specifications for $f_p^\prime=0.2$~Hz, $f_p^{\prime\prime}=1$~Hz, and $f_p^{\prime\prime\prime}=5$~Hz. In contrast, the GFL model lies outside our bounds.

To further support these observations, a loss of last synchronous machine has been simulated for a plant model containing the aforementioned IBRs. As expected the GFM model survives the loss of the last synchronous machine while the GFL model does not. Notably, both IBR models supply $90\%$ of the lost generation within one cycle, highlighting that this commonly used criterion is insufficient to distinguish between GFM and GFL responses.

\subsection{Battery Energy Storage System (OEM 2)}
\begin{figure}[b!!]
    \centering
    \vspace{-1em}
\includegraphics[width=1\columnwidth]{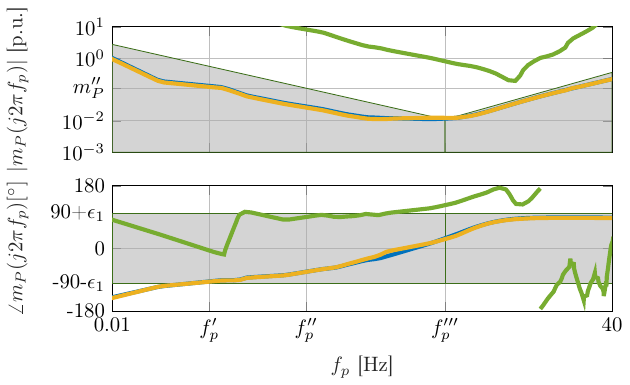}
\vspace{-2em}
\caption{Dynamic droop coefficient $m_P( j 2 \pi f_p)$ for GFL (\drawline[mycolor3, line width=1.3pt]), GFM~A (\drawline[mycolor1, line width=1.3pt]), and GFM~B (\drawline[mycolor2, line width=1.3pt]) configuration of the model of OEM~2.\label{fig:OEM2_mp}}
\vspace{0.5em}
\includegraphics[width=1\columnwidth]{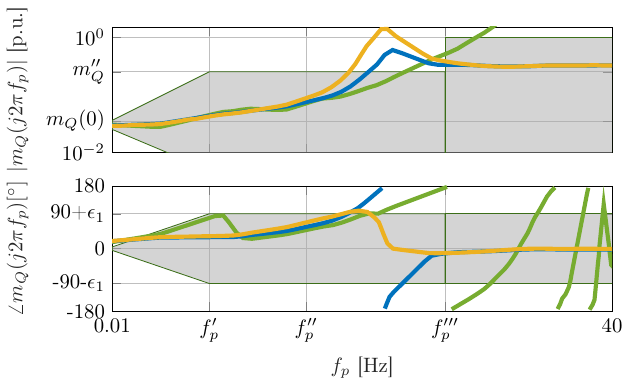}
\vspace{-2em}
\caption{Dynamic droop coefficient $m_Q( j 2 \pi f_p)$ for GFL (\drawline[mycolor3, line width=1.3pt]), GFM~A (\drawline[mycolor1, line width=1.3pt]), and GFM~B (\drawline[mycolor2, line width=1.3pt]) configuration of the model of OEM~2.\label{fig:OEM2_mq}}
\end{figure}
\begin{figure}[t!!]
    \centering
\includegraphics[width=\columnwidth]{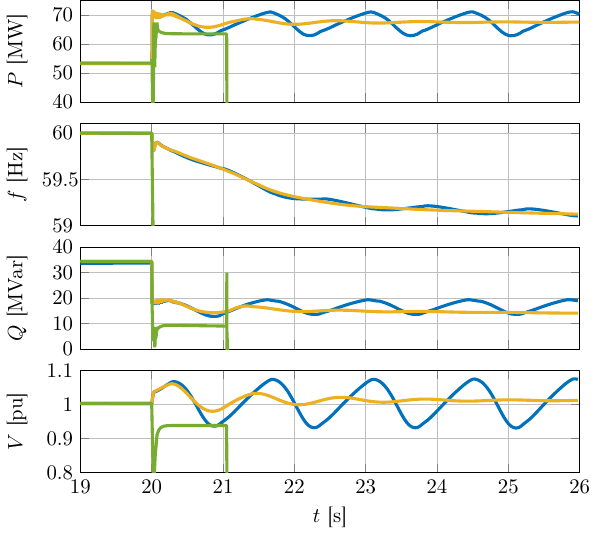}
        \vspace{-1em}
\caption{Loss of last synchronous machine for the OEM 2 BESS in GFL (\drawline[mycolor3, line width=1.3pt]),  GFM A (\drawline[mycolor1, line width=1.3pt]), and GFM B (\drawline[mycolor2, line width=1.3pt]) configurations. The GFL IBR (green) becomes unstable after $t=21$~s.\label{fig:OEM2_loss}}
\end{figure}

Finally, we scan a $107$~MVA, $220$~kV IBR plant model provided by a vendor (OEM~2). The plant model can be configured to use either GFL control or one of two GFM controls. In all configurations, for the purpose of this study, steady-state frequency droop is turned off and the plant operates in a voltage control mode instead of $Q\text{-}V$ droop. Bode plots of the dynamic droop coefficients $m_P( j 2 \pi f_p)$ and $m_Q( j 2 \pi f_p)$ are shown in Fig.~\ref{fig:OEM2_mp} and  Fig.~\ref{fig:OEM2_mq}. The configuration GFM~B largely aligns with the proposed specifications for $f_p^\prime=0.05$~Hz, $f_p^{\prime\prime}=0.25$~Hz, and $f_p^{\prime\prime\prime}=2.5$~Hz, but exhibits violations of the gain bound on $m_Q( j 2 \pi f_p)$ due to insufficient voltage control bandwidth. In other words, for $f_p \in [f_p^{\prime\prime},f_p^{\prime\prime\prime}]$ the voltage magnitude is no longer stiffly controlled. However, with $\angle m_Q( j 2 \pi f_p) \approx 90^\circ$, the IBR plant merely exhibits incorrect gain and no significant negative damping. While the configuration GFM~A broadly exhibits similar dynamics, it significantly violates the phase bound on $m_Q( j 2 \pi f_p)$ for $f_p \in [0.4,1.5]$~Hz and shows a $180^\circ$ phase shift at $f_p \approx 0.8$~Hz, i.e., exhibits significant negative damping. Finally, the GFL configuration largely lies outside our specifications.
The results of a loss of last synchronous machine test are shown in Fig.~\ref{fig:OEM2_loss}. Again, all models provide $90\%$ of the active power deficit within one cycle of the event. As expected, the GFL configuration does not survive the event. Moreover, GFM~A exhibits a significant $0.8$~Hz oscillation that can be explained by the $180^\circ$ phase shift of $m_Q( j 2 \pi f_p)$ at around $0.8$~Hz. Finally, we note that, due to the lack of steady-state frequency droop the frequency in Fig.~\ref{fig:OEM2_loss} is slowly diverging from $\omega_0$, highlighting the proportional-integral nature of $m_P( j 2 \pi f_p)$ as $f_p \to 0$.

\section{Conclusion}
This work proposed to use dynamic droop coefficients to specify functional capabilities of GFM IBRs resources and complement widely considered time domain tests and impedance specifications. To this end, we developed minimum requirements for GFM IBRs that aim to rule out adverse interactions. Moreover, we developed bounds on the dynamic droop coefficients that, in abstraction, encode typical voltage and frequency control functions, some of which may also be provided by GFL IBRs. EMT simulations of common GFL and GFM controls as well as OEM models are used to illustrate the results and highlight the usefulness of the dynamic droop coefficients and specifications both as a screening tool. Determining bounds on the dynamic droop coefficients to ensure small-signal stability while fully accounting for the impact of cross coupling are seen as an important topic for future work. 



\bibliographystyle{ieeetr}
\bibliography{IEEEabrv,references}

\appendices

\balance

\end{document}